\documentclass{article}
\usepackage[preprint]{neurips_2023}
\usepackage{graphicx}

\usepackage[utf8]{inputenc} 
\usepackage[T1]{fontenc}    
\usepackage{hyperref}       
\usepackage{url}            
\usepackage{booktabs}       
\usepackage{amsfonts}       
\usepackage{nicefrac}       
\usepackage{microtype}      
\usepackage{xcolor}         
\usepackage{tablefootnote}
\usepackage{amsmath, bm}
\usepackage{algorithm}
\usepackage{algpseudocode}

\begin{document}

\def\x{{\mathbf x}}
\def\L{{\cal L}}

\title{CoMoSVC: Consistency Model-based Singing Voice Conversion}
\author{Yiwen Lu\textsuperscript{1}, Zhen Ye\textsuperscript{1}, Wei Xue\textsuperscript{1$\dag$}, Xu Tan\textsuperscript{2}, Qifeng Liu\textsuperscript{1}, Yike Guo\textsuperscript{1$\dag$} 
\\
\textsuperscript{1}	Hong Kong University of Science and Technology \textsuperscript{2} Microsoft Research Asia 
\thanks{$^\dag$Corresponding authors: Wei Xue \{\small weixue@ust.hk\}, Yike Guo \{\small yikeguo@ust.hk\}        
}
}

\maketitle

\begin{abstract}
The diffusion-based Singing Voice Conversion (SVC) methods have achieved remarkable performances, producing natural audios with high similarity to the target timbre. However, the iterative sampling process results in slow inference speed, and acceleration thus becomes crucial. In this paper, we propose CoMoSVC, a consistency model-based SVC method, which aims to achieve both high-quality generation and high-speed sampling. A diffusion-based teacher model is first specially designed for SVC, and a student model is further distilled under self-consistency properties to achieve one-step sampling. Experiments on a single NVIDIA GTX4090 GPU reveal that although CoMoSVC has a significantly faster inference speed than the state-of-the-art (SOTA) diffusion-based SVC system, it still achieves comparable or superior conversion performance based on both subjective and objective metrics. Audio samples and codes are available at \href{https://comosvc.github.io/}{https://comosvc.github.io/}.

\end{abstract}
{\bfseries Keywords:} {Singing Voice Conversion, Diffusion Model, Consistency Model}

\section{Introduction}
\label{sec:intro}

Singing Voice Conversion(SVC) aims to convert one singer's voice to another one's, while preserving the content and melody. It has wide applications in music entertainment, singing voice beautification, and art creation  \cite{zhang2023leveraging}. 

Statistical methods  \cite{kobayashi2014statistical,kobayashi2015statistical,kobayashi2015statistical1} are applied to the SVC tasks with parallel training data from both the source and target singers, which is usually infeasible, and thus the non-parallel SVC methods have become the mainstream. Two-stage methods are generally used for SVC, the first stage disentangles and encodes singer-independent and singer-dependent features from the audio. Then the second decoding stage generates the converted audio by replacing the singer-dependent feature with the target one. Since the substantial impact of the second stage on the quality of the converted audios, it has become crucial to design and optimize this stage. Therefore, many generative models have been used for the SVC decoding, including the autoregressive (AR) models, generative adversarial network (GAN), Normalizing Flow, and diffusion models. AR models are firstly used to develop USVC  \cite{nachmani2019unsupervised}, and PitchNet  \cite{deng2020pitchnet} further improves USVC by adding a pitch adversarial network to learn the joint phonetic and pitch representation. However, AR models are slow due to the recursive nature, then non-AR GAN-based UCD-SVC  \cite{polyak2020unsupervised} and FastSVC  \cite{liu2021fastsvc} are later proposed. Since the unstable training of GAN, a flow-based end-to-end SVC system, named SoVITS-SVC  \cite{SVC2023} received widespread attention for its excellent converted results in fast speed. Recently, it has been shown that the conversion performance can be substantially improved by the diffusion-based SVC methods such as DiffSVC  \cite{liu2021diffsvc} and the diffusion version of SoVITS-SVC.

However, the iterative sampling process results to the slow inference of the diffusion-based SVC methods. A new generative model named consistency model  \cite{song2023consistency} has been proposed to realize one-step generation. Subsequently for speech synthesis, CoMoSpeech  \cite{ye2023comospeech} exploits the consistency model to achieve both high-quality synthesis and fast inference speed. Inspired by this, a consistency model-based SVC method, named CoMoSVC, is further developed in this paper to achieve {\em high-quality, high-similarity and high-speed} SVC. Based on the structure of EDM  \cite{karras2022elucidating}, a diffusion-based teacher model with outstanding generative capability is firstly designed, and a student model is further distilled from it to achieve one-step sampling. Experiments reveal that while the sampling speed of CoMoSVC is approximately 500 and 50 times faster than that of the diffusion-based SoVITS-SVC and DiffSVC respectively, the comparable performance is still retained and some improvements can even be achieved in both quality and similarity.

\section{Background}

The diffusion model generates samples by first adding noise to data during the forward process and then reconstructing the data structure in the reverse process.
We assume that the original data distribution is $p_{data}(\mathbf{x})$, and the forward process can be represented by a stochastic differential equation (SDE)  \cite{song2020score,karras2022elucidating}:
\begin{equation}
\mathrm{d}\mathbf{x}_t=\mathbf{f}(\mathbf{x}_t ,t)\mathrm{d}t+g(t)\mathrm{d}\mathbf{w}_t ,
\end{equation}
where $\mathbf{w}_t$ is the standard wiener process, $\mathbf{f}(\mathbf{x}_t ,t)$ and $g(t)$ are drift and diffusion coefficients respectively. With setting $\mathbf{f}(\mathbf{x}_t,t)=0$ and $g(t)=\sqrt{2t}$, the same as the choice in  \cite{karras2022elucidating}, the SDE can be defined by:
\begin{equation}
\mathrm{d}\mathbf{x}_{t}=\sqrt{2t}\,\mathrm{d}\mathbf{w} _{t}. 
\end{equation}
The reverse process can also be expressed by a reverse-time SDE  \cite{song2020score}:
\begin{equation}
\mathrm{d}\mathbf{x}_{t}=-2t\nabla\mathrm{log}\,p_t(\mathbf{x}_t)dt+\sqrt{2t}\,\mathrm{d}\bar{\mathbf{w}}_{t} ,
\end{equation}
where $p_t(\mathbf{x}_t)$ is the distribution of $\mathbf{x}_t$, $\nabla\mathrm{log}\,p_t(\mathbf{x}_t)$ is the score function, and $\bar{\mathbf{w}}_{t}$ is the reverse-time standard wiener process.  \cite{song2020score} found that there exists a probability flow (PF) ordinary differential equation (ODE), whose solution trajectories' distribution at time $t$ is the same as $p_t(\mathbf{x}_t)$. The PF ODE with such property can be represented by
\begin{equation}
\frac{\mathrm{d} \mathbf{x}_t}{\mathrm{d} t}=-t \nabla \log p_t\left(\mathbf{x}_t\right)=\frac{\mathbf{x}_t-D_\phi\left(\mathbf{x}_t, t\right)}{t},
\label{4}
\end{equation} 
where $D_\phi$ is the neural network with $\phi$ as parameters to approximate the denoiser function. Then for sampling, the PF ODE is solved by initiating from $\mathbf{x}_T$, as
\begin{equation}    \mathbf{x}_0=\mathbf{x}_T+\int_T^0 \frac{\mathbf{x}_t-D_\phi\left(\mathbf{x}_t, t\right)}{t} \mathrm{~d} t.
\label{5}
\end{equation}
However, the diffusion model generally needs a large number of iterations to solve the PF ODE, making the sampling slow. A consistency model \cite{song2023consistency} is proposed for one-step sampling based on the self-consistency property, making any point from the same PF ODE trajectory be mapped to the same initial point. The self-consistency properties have two constraints: firstly, any pair of points $\mathbf{x}_{t_m}$ and $\mathbf{x}_{t_n}$ will be mapped to the same point, which can be represented by:
\begin{equation}    D_\phi(\mathbf{x}_{t_m},t_m)=D_\phi(\mathbf{x}_{t_n},t_n).
\end{equation}
Secondly, the initial point should also be mapped to itself and this constraint is called the boundary condition. To avoid numerical instability, it can be given by
\begin{equation}
    D_\phi(\mathbf{x}_\epsilon,t_\epsilon)=\mathbf{x}_\epsilon,
\label{7}
\end{equation}
where $\epsilon$ is a fixed small positive number and set as 0.002.
, all the singers have their identification number, which will be encoded as a singer embedding. 

\section{Proposed Method}
CoMoSVC is a two-stage model, where the first stage encodes the extracted features and the singer identity into embeddings. These embeddings are concatenated and serve as the conditional input for the second stage to generate mel-spectrogram, which can be further rendered to audio by using a pre-trained vocoder. The training process depicted in Fig.~\ref{a_training} takes the waveform and its singer identity as the input to reconstruct the mel-spectrogram, while the inference process illustrated in Fig.~\ref{a_como}  replaces the singer identity with the target one to generate the converted mel-spectrogram.

\begin{figure}[t]
      \centering
      \includegraphics[width=\linewidth]{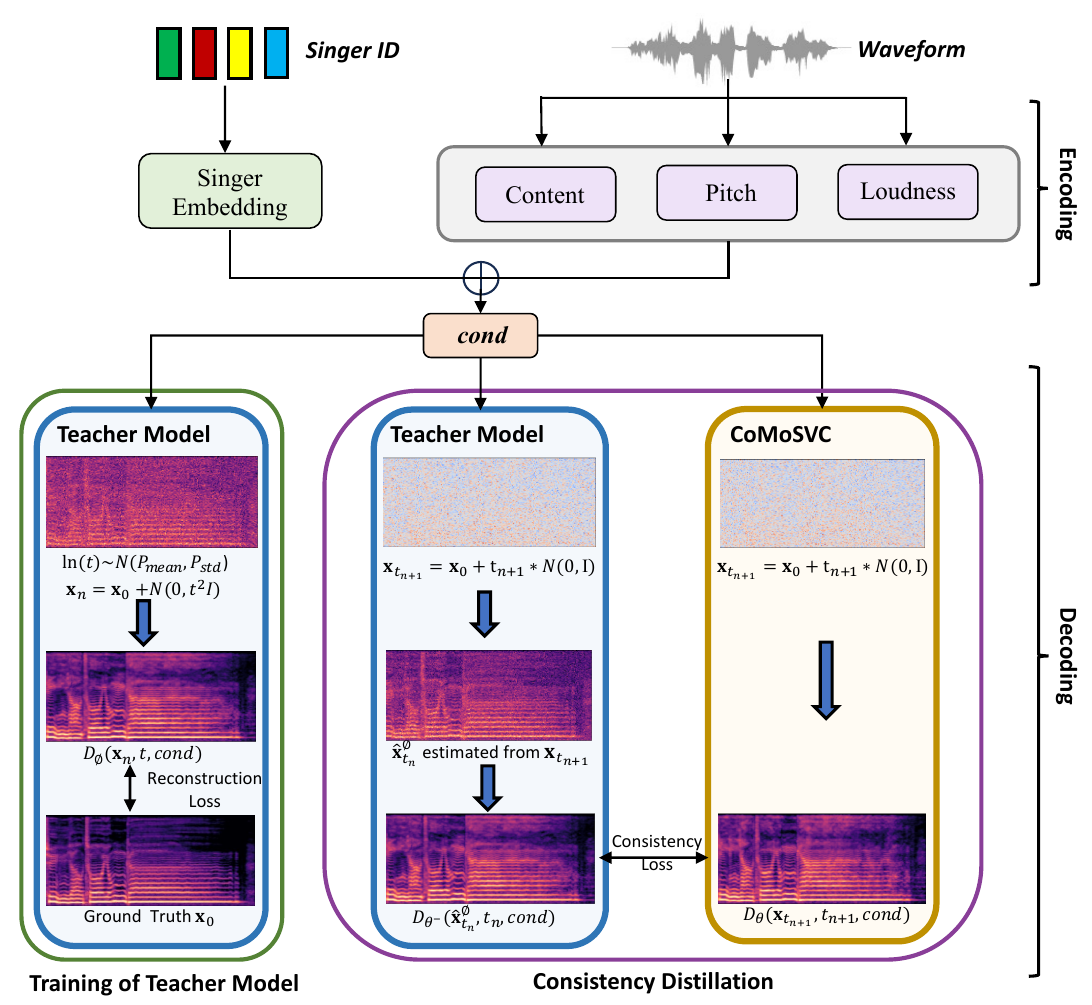}
      \caption{The Training Process.}
      \label{fig1}
\end{figure}

\begin{figure}[!ht]
      \centering
      \includegraphics[width=\linewidth]{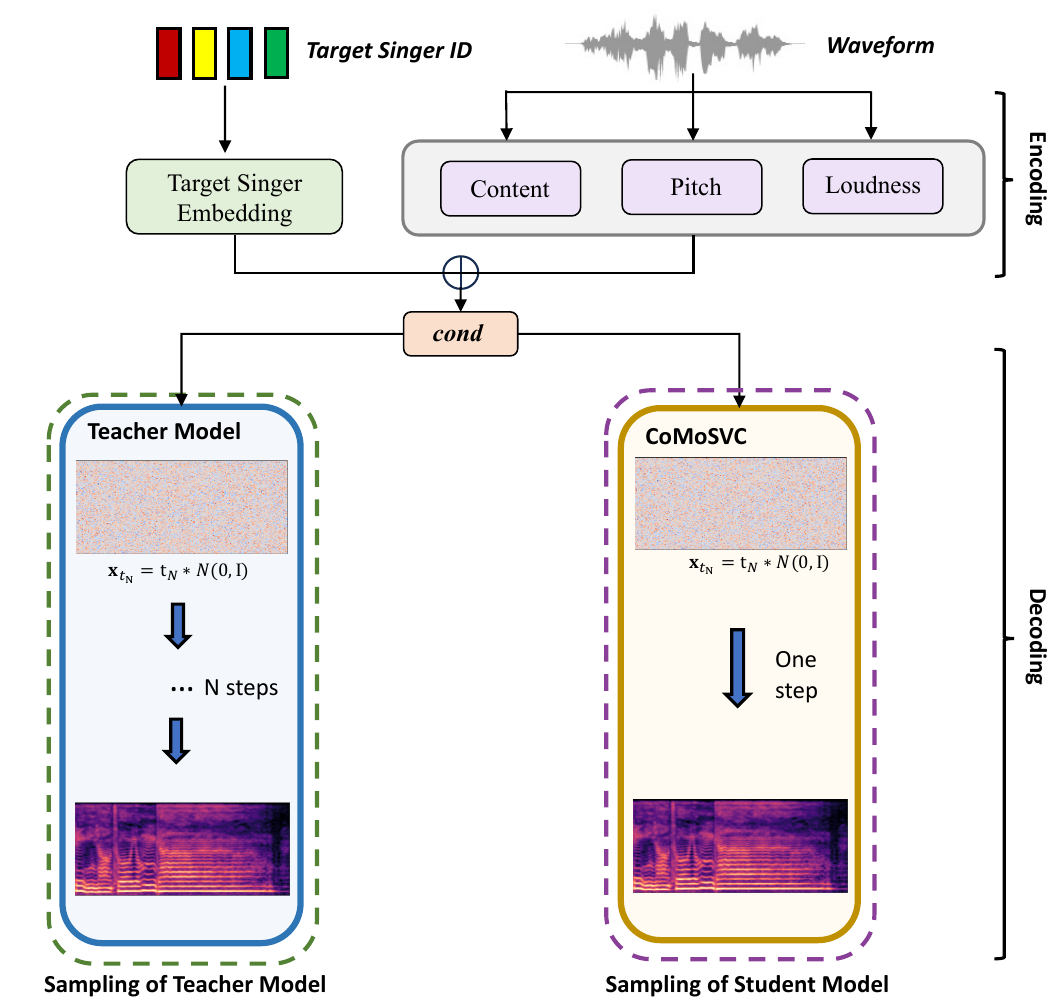}
      \caption{The inference process.}
      \label{fig2}
\end{figure}

\subsection{Encoding}
This section encodes both singer-independent and singer-dependent features, which can be shown in the upper part of both Fig.~\ref{fig1} and Fig.~\ref{fig2}. We extract content, pitch, and loudness features to capture singer-independent information in audio, while the singer ID is used as the singer-dependent information. The content features are extracted by using the pre-trained acoustic model ContentVec  \cite{qian2022contentvec} and the large dimensionality of these features allows for enhancing the clarity of lyrics in the converted audio. To represent pitch information, we use the widely-used and classical F0 estimator DIO  \cite{morise2009fast}. The squared magnitude of the audio signal is calculated as the loudness feature. After feature extraction, we applied a linear layer to all the embeddings to unify the dimensions and concatenate them to form the conditional input for the decoding stage.

\subsection{Decoding}
This stage is the key component of CoMoSVC, during which the mel-spectrograms can be generated from the conditional input. A teacher model is first trained and then a student model is distilled from it, which will be introduced in section.~3.2.1 and section.~3.2.2 respectively. The sampling process of both the teacher model and student model will be explained in section.~3.2.3.

\subsubsection{Teacher Model}
We use the architecture of EDM  \cite{karras2022elucidating} as the teacher model to train the denoiser function $D_\phi$ due to its high generative ability. Moreover, the structure of $D_\phi$ used here is the non-causal Wavenet  \cite{rethage2018wavenet}. We use $\mathbf{x}_0\sim p_{data}(\mathbf{x})$ and $\mathit{cond}$ to denote the ground truth mel-spectrogram and the conditional input. According to \eqref{4}, the empirical ODE is 
\begin{equation}
        \frac{\mathrm{d} \mathbf{x}_t}{\mathrm{d} t}=\frac{\mathbf{x}_t-D_\phi\left(\mathbf{x}_t, t,\mathit{cond}\right)}{t},
\end{equation}
where $\mathbf{x}_t=\mathbf{x}_0+t*\mathcal{N}(0,\boldsymbol{I})$, represents the result after adding noise. Similar to  \cite{karras2022elucidating}, we use a different network $F_\phi$ instead of directly approximating the denoiser function by $D_\phi$. The network is preconditioned with a skip connection to make the estimation more flexible, it can be given by
\begin{equation}
    D_\phi(\mathbf{x}_t,t,\mathit{cond})=c_{\text {skip }}(t) \mathbf{x}_t+c_{\text {out}}(t) F_\phi\left(c_{\text {in }}(t) \mathbf{x}_t,t, c_{\text {noise }}(t)\right).
\end{equation}
$c_{\text {skip}}(t)$ modulates the skip connection, $c_{\text {in }}(t)$ and $c_{\text {out }}(t)$ scale the magnitudes of $\mathbf{x}_t$ and $F_\phi$ respectively, and $c_{\text {noise }}(t)$ maps noise level $t$ into a conditioning input for $F_\phi$.
To satisfy the boundary condition mentioned in \eqref{7} and ensure $c_{\text {skip }}(t)$ and $c_{\text {out }}(t)$ differential, we choose
\begin{equation}
    c_{\text {skip }}(t)=\frac{\sigma_{\text {data }}^2}{(t-\epsilon)^2+\sigma_{\text {data }}^2}, \quad c_{\text {out }}(t)=\frac{\sigma_{\text {data }}(t-\epsilon)}{\sqrt{\sigma_{\text {data }}^2+t^2}} ,
\end{equation}
where $\sigma_{\text {data }}$ is the standard deviation of $p_{data}(\mathbf{x})$.
The loss function $\mathcal{L}_\phi$ used to train the $D_\phi$ can be designed by
\begin{equation}
    \mathcal{L}_\phi=\mathbb{E}[\lambda(t)\| D_\phi\left(\mathbf{x}_t, t, cond\right)-\mathbf{x}_0 \|^2] ,
\end{equation}
where $\lambda(t)=(t^2+\sigma_{\text {data }}^2)/(t\cdot\sigma_{\text {data }} )^2$, denotes the weight corresponding to different noise level $t$. The entire procedure is depicted in the lower left section in Fig.~\ref{fig1}.

\begin{algorithm}[t]
\caption{Training procedure}\label{a_training}
\begin{algorithmic}[1]
\Statex \textbf{Input:}  The denoiser function $D_\phi$ of the teacher model; the conditional input $cond$; the original data distribution $p_{data}$ and $\mu$
\State \textbf{repeat}
\State Sample $n \sim \mathcal{U}(1,N-1)$ and $\mathbf{x}_0 \sim p_{data}$
\State Sample $\mathbf{x}_{t_{n+1}} \sim \mathcal{N}(\mathbf{x}_0,(t_{n+1})^2*I)$ 
\State $\hat{\mathbf{x}}_{t_n}^\phi \leftarrow \frac{t_n}{t_{n+1}}\mathbf{x}_{t_{n+1}}+ \frac{t_{n+1}-t_n}{t_{n+1}}D_\phi\left(\mathbf{x}_{t_{n+1}}, t_{n+1},cond\right)$
\State $\mathcal{L}_\theta\leftarrow d\left(D_\theta\left(\mathbf{x}_{t_{n+1}}, t_{n+1},cond\right),D_{\theta^{-}}\left(\hat{\mathbf{x}}_{t_n}^\phi, t_n, cond\right)\right)$
\State $\boldsymbol{\theta} \leftarrow \boldsymbol{\theta}-\eta \nabla_{\boldsymbol{\theta}} \mathcal{L}\left(\boldsymbol{\theta},\boldsymbol{\theta}-;\boldsymbol{\phi}\right)$
\State $\boldsymbol{\theta}^{-} \leftarrow \operatorname{stopgrad}\left(\mu \boldsymbol{\theta}^{-}+(1-\mu) \boldsymbol{\theta}\right)$
\State \textbf{until} convergence

\end{algorithmic}
\end{algorithm}

\begin{algorithm}[t]
\caption{Sampling procedure}\label{a_como}
\begin{algorithmic}[1]
\Statex \textbf{Input:}  The denoiser function $D_\theta$ of the consistency model; the conditional input $cond$ ; a set of time points $t_{i\in\{0,\ldots,N\}}$
\State Sample $\mathbf{x}_{N} \sim \mathcal{N}(0,\sigma(t_N)^{2}*I)$
\State  $\mathbf{x} \leftarrow D_{\theta}(\mathbf{x}_{N},t_{N},cond)$ 
\State  \textbf{if}  one-step synthesis
\State \hskip1.0em \textbf{Output:} $\mathbf{x}$
\State  \textbf{else}            multi-step synthesis
\State \hskip1.0em \textbf{for} $i = N-1$ \textbf{to} $1$ \textbf{do}                
\State \hskip1.0em  \hskip1.0em  Sample $\textbf{z} \sim \mathcal{N}(0,\boldsymbol{I})$
\State \hskip1.0em \hskip1.0em  $\mathbf{x}_{i} \leftarrow  \mathbf{x} +\sqrt{t_{i}^{2}-\epsilon^{2}}{\bf z} $
\State \hskip1.0em \hskip1.0em  $\mathbf{x} \leftarrow D_{\theta}(\mathbf{x}_{i},t_{i},cond)$                       
\State \hskip1.0em \textbf{end for}
\Statex \hskip1.0em \textbf{Output:} $\mathbf{x}$
 
\end{algorithmic}
\end{algorithm}

\subsubsection{Consistency Distillation}
A student model can be further distilled from the pre-trained denoiser function $D_\phi$ to ultimately achieve one-step sampling, the process is illustrated in Algorithm.~\ref{a_training} and the lower right section of Fig.~\ref{a_training}. First, we randomly sample $n$ from the uniform distribution $\mathcal{U}(1,N-1)$ and obtain $\mathbf{x}_{t_{n+1}}$ by adding $t_{n+1}*\mathcal{N}(0,\boldsymbol{I})$ to $\mathbf{x}_0$, then we use the $D_\phi$ to get the one-step estimation $\hat{\mathbf{x}}_{t_n}^{\phi }$ from $\mathbf{x}_{t_{n+1}}$. According to \eqref{4}, since first-order Euler Solver is used here, it can be given by
\begin{equation}
    \hat{\mathbf{x}}_{t_n}^{\phi } =\frac{t_n}{t_{n+1}}\mathbf{x}_{t_{n+1}}+ \frac{t_{n+1}-t_n}{t_{n+1}}D_\phi\left(\mathbf{x}_{t_{n+1}}, t_{n+1},cond\right) .
\end{equation}
The structure of the student model is inherited from the teacher model's denoiser function $D_\phi$, resulting in $D_\theta$ and $D_{\theta^{-}}$. The parameters $\theta$ and $\theta^{-}$ are initialized with $\phi$, $\theta^{-}$ is a running average of the past values of $\theta$. Afterwards, we use $D_{\theta^{-}}\left(\hat{\mathbf{x}}_{t_n}^\phi, t_n, cond\right)$ and $D_\theta\left(\mathbf{x}_{t_{n+1}}, t_{n+1},cond\right)$ to obtain different outputs of the pair of adjacent points $\hat{\mathbf{x}}_{t_n}^{\phi }$ and $\mathbf{x}_{t_n}$. The consistency distillation is trained by minimizing the $L_{2}$  distance between the two outputs: 
\begin{equation}
\begin{aligned}
     &d\left(D_\theta\left(\mathbf{x}_{t_{n+1}}, t_{n+1},cond\right),D_{\theta^{-}}\left(\hat{\mathbf{x}}_{t_n}^\phi, t_n, cond\right)\right)\\
     &=\| D_\theta\left(\mathbf{x}_{t_{n+1}}, t_{n+1},cond\right)-D_{\theta^{-}}\left(\hat{\mathbf{x}}_{t_n}^\phi, t_n, cond\right) \|^2.
\end{aligned}
\end{equation}

The parameter $\theta$ is updated by:
\begin{equation}
    \boldsymbol{\theta} \leftarrow\boldsymbol{\theta}-\eta \nabla_{\boldsymbol{\theta}} \mathcal{L}\left(\boldsymbol{\theta},\boldsymbol{\theta}-;\boldsymbol{\phi}\right).
\end{equation}
To stabilize the training, the exponential moving average (EMA) update and stop grad are adopted to $\theta^{-}$, as:
\begin{equation}
    \boldsymbol{\theta}^{-} \leftarrow \operatorname{stopgrad}\left(\mu \boldsymbol{\theta}^{-}+(1-\mu) \boldsymbol{\theta}\right),
\end{equation}
where $\mu$ is a momentum coefficient, empirically set as 0.95.

\subsubsection{Sampling Process}
The sampling processes of both the two models are depicted in the lower part of Fig.~\ref{fig2}. The teacher model takes a number of iterations for sampling, while the student model can achieve one-step sampling as summarized in Algorithm.~\ref{a_como}. We first sample the noise that has the same shape as the mel-spectrogram by $\mathbf{x}_{t_N}=t_N*\mathcal{N}(0,\boldsymbol{I})$, and the output of $D_\theta(\mathbf{x}_{t_{N}},t_N,cond)$ is the sampling result. The proposed CoMoSVC also supports multi-step sampling by chaining the outputs at multiple time steps. However, there will be a trade-off between the number of iterations and sampling quality.

\section{Experiments}

\subsection{Experimental Setup}
We conduct our experiment on two open-source datasets, which are M4Singer  \cite{zhang2022m4singer} and OpenSinger  \cite{huang2021multi}, respectively. The former dataset has 29.77 hours of singing voice and 20 singers, and the latter one contains 50 hours and 66 singers. All the audios are resampled to 24kHz and normalized. Then we calculate the volume feature, extract the F0 curve along with the voiced/unvoiced flag for each frame by using DIO  \cite{morise2009fast} and the 768-dimensional content feature from the 12th layer by utilizing ContentVec \cite{qian2022contentvec}. All these features are projected to 256 dimensions and then concatenated as the conditional input for the decoding stage. We use the vocoder\footnote{https://github.com/M4Singer/M4Singer/tree/master/code} pre-trained with singing voice from M4singer \cite{zhang2022m4singer}, and the mel-spectrograms are computed with 512-point fast Fourier transform (FFT), 512-point window size and 128-point hop size with 80 frequency bins. 

All the models are trained for 1 million iterations on a single NVIDIA GTX4090 GPU with a batch size of 48, with learning rates as 1e-4 and 5e-5 respectively and the optimizer is AdamW. 

We first conduct the reconstruction experiment to evaluate the capabilities of different decoding stages in the autoencoding settings. Then two sets of experiments are conducted for any-to-many SVC task, a) train on the OpenSinger dataset for the target singer and use M4singer as the source for conversion ; b) train on the M4singer dataset for the target singer and use OpenSinger as the source for conversion. Moreover, we increase the sampling steps of CoMoSVC and conduct the conversion experiments to evaluate the effect of sampling steps.

\subsection{Baselines}

We compare the proposed CoMoSVC with the SOTA SVC methods, including: 

\begin{itemize}
\item SoVITS-Flow: The flow version of SoVITS-SVC\footnote{https://github.com/svc-develop-team/so-vits-svc?tab=readme-ov-file\#sovits-model}.
\item SoVITS-Diff: The diffusion version of SoVITS-SVC\footnote{{https://github.com/svc-develop-team/so-vits-svc?tab=readme-ov-file\#diffusion-model-optional}}. The number of diffusion steps is 1000.
\item DiffSVC \cite{liu2021diffsvc}: The first SVC method based on the diffusion model, and the number of steps is 100.
\end{itemize}
The same feature embeddings and training details as described in Sec.~4.1 are used for the baseline methods for fair comparison.

\subsection{Evaluation}
We evaluate the reconstruction ability of different methods by objective metrics and the conversion ability by both subjective and objective metrics. For the subjective test, we invited 12 volunteers to give Mean Opinion Score (MOS) on naturalness and similarity on the converted audios. Real-Time Factor (RTF), Character Error Rate (CER) obtained by Whisper \cite{radford2023robust} and speaker SIMilarity (SIM) calculated by the cosine distance between the speaker embeddings \footnote{https://github.com/Jungjee/RawNet} are used as the objective metrics for SVC evaluation. Since the flow version SoVITS-SVC is end-to-end and the other methods are two-stage, we use the time ratio of transforming embeddings into latent representations (Flow) / mel-spectrograms (Others) to the duration of audio to represent RTF for clear comparison. 
For reconstruction, F0 Pearson correlation coefficient (FPC) \footnote{{https://github.com/open-mmlab/Amphion/blob/main/evaluation/metrics/f0/}} and PESQ \cite{ITUT2001} are used additionally for evaluation.

\begin{table}[t]
    \begin{center}
    \caption{Objective Evaluations for Reconstruction}
    \label{table1}
    {  
    \begin{tabular}{@{}lccccc@{}}
    \toprule
    METHOD 	&NFE($\downarrow$) &FPC($\uparrow$)	 &PESQ($\uparrow$)	&CER($\downarrow$) &SIM($\uparrow$) \\ \toprule
    SoVITS-Flow	 &1 &0.935	&2.486 &25.03	&0.948\\
    SoVITS-Diff	 &1000 &0.938	&2.826 &26.04	&0.970\\
    DiffSVC	 &100 &0.941	&2.917 &25.47	&0.972 \\\midrule
    Teacher	&50 &\textbf{0.945}	&\textbf{2.967}	&\textbf{20.96}	&\textbf{0.982}\\
    \textbf{CoMoSVC}	&\textbf{1} &0.943	&2.948	&24.69	&0.970\\ \bottomrule
    \end{tabular}}
    \end{center}
    \end{table}
    
    \begin{table}[t]
    \begin{center}
    \caption{Objective Evaluations for SVC, where ``M'' and ``O'' in the SVC setting row stands for M4Singer and OpenSinger, respectively.}
    \label{table2}
    {  
    \begin{tabular}{@{}lcccccc@{}}
    \toprule
    SVC SETTING&\multicolumn{1}{c}{}&\multicolumn{2}{c}{M$\to$O}&\multicolumn{2}{c}{O$\to$M}\\\midrule 
    METHOD &NFE($\downarrow$) &SIM($\uparrow$) &CER($\downarrow$) &SIM($\uparrow$) &CER($\downarrow$) \\\toprule
    SoVITS-Flow	&1 &0.784	&21.76 &0.585	&22.62\\ 
    SoVITS-Diff	 &1000 &\textbf{0.804}	&\textbf{20.40} &0.598	&25.50\\ 
    DiffSVC	  &100	&0.801 	&21.27 &0.598 	&22.58 \\ \midrule
    Teacher	&50	&0.794 	&20.52 &\textbf{0.614} 	&\textbf{19.57} \\ 
    \textbf{CoMoSVC}	&\textbf{1} &0.801 	&21.49 &0.585 	&19.76\\ \bottomrule
    
    \end{tabular}}
    \end{center}
    \end{table}

\subsubsection{Reconstruction}
As illustrated in Table.~\ref{table1}, the teacher model outperforms all the models in all the metrics. Similarly, CoMoSVC outperforms all the baselines in all metrics except for similarity, where it achieves comparable results. This indicates the outstanding generative ability of the decoding stage of CoMoSVC with only one step, which is hundreds or thousands of times fewer than all the baselines.
\subsubsection{SVC Performances}
We conduct two sets of SVC experiments as described in section.~4.1 and the source audios are all unseen during training. As illustrated by CER and SIM in Table.~\ref{table2}, CoMoSVC performs comparably to all the baselines. The subjective evaluations in Table.~\ref{table3} reveal that CoMoSVC achieves comparable naturalness to the diffusion-based SVC methods. Furthermore, the similarity of CoMoSVC exceeds that of all the baselines in both experiments, demonstrating an improvement of at least 0.05 to diffusion-based SVC methods. Moreover, both the naturalness and similarity of CoMoSVC show an increment of approximately 1 compared to the flow version SoVITS-SVC. As to the inference speed, the RTF of CoMoSVC is 0.002 smaller than that of the flow version SoVITS-SVC. In comparison with the diffusion-based SVC methods, CoMoSVC is more than 45 times faster than DiffSVC and almost 500 times faster than the diffusion-version SoVITS-SVC.

\begin{table}[t]
    \centering
    \caption{Subjective Evaluations for SVC, where, as an example, M4Singer$\to$OpenSinger means converting the singing voices from the M4Singer to target timbres in the OpenSinger.}
    \label{table3}
    \begin{tabular}{@{}lcccccc@{}}
    \toprule
    SVC SETTING&\multicolumn{2}{c}{}&\multicolumn{2}{c}{M4Singer$\to$OpenSinger}&\multicolumn{2}{c}{OpenSinger$\to$M4Singer}\\\midrule 
    METHOD &NFE($\downarrow$)  &RTF($\downarrow$)	&MOS/N($\uparrow$) &MOS/S($\uparrow$) &MOS/N($\uparrow$) &MOS/S($\uparrow$)\\\toprule
    SoVITS-Flow	&1 &0.008   &3.27 $\pm$ 0.21 &3.03 $\pm$ 0.23 	&3.10 $\pm$ 0.22 	&2.90 $\pm$ 0.23 \\
    SoVITS-Diff	&1000 &2.978  &4.43 $\pm$ 0.14 &3.90 $\pm$ 0.19 	&4.32 $\pm$ 0.15 	&3.99 $\pm$ 0.19 \\
    DiffSVC	&100 &0.278  &\textbf{4.44 $\pm$ 0.14} 	&3.91 $\pm$ 0.19 	&4.23 $\pm$ 0.19 	&3.95 $\pm$ 0.21 \\\midrule
    Teacher &50	 &0.148   &4.43 $\pm$ 0.15 	&3.92 $\pm$ 0.19 	&\textbf{4.47 $\pm$ 0.14} 	&\textbf{4.05 $\pm$ 0.18} \\
    \textbf{CoMoSVC}  &\textbf{1}	&\textbf{0.006  } &4.42 $\pm$ 0.13 	&\textbf{3.96 $\pm$ 0.19} 	&4.27 $\pm$ 0.16 	&4.00 $\pm$ 0.19 \\ \bottomrule
    \end{tabular}
    \end{table}

\begin{table}[t]
\centering
\caption{Evaluations for Effect of Sampling Steps, where the number in the method name represents the number of sampling steps.}
\label{table4}
\begin{tabular}{@{}lcc@{}}
\toprule
SVC SETTING&\multicolumn{2}{c}{M$\to$O}\\\midrule 
METHOD 	&MOS/N($\uparrow$) &MOS/S($\uparrow$) \\\toprule
CoMoSVC-4   &\textbf{4.46 $\pm$ 0.14} 	&3.95 $\pm$ 0.19 	\\
CoMoSVC-2  &4.36 $\pm$ 0.14 	&3.88 $\pm$ 0.19 	\\
\textbf{CoMoSVC-1}  &4.42 $\pm$ 0.13 	&\textbf{3.96 $\pm$ 0.19} 	\\ \midrule 
SVC SETTING&\multicolumn{2}{c}{O$\to$M}\\\midrule 
METHOD 	&MOS/N($\uparrow$) &MOS/S($\uparrow$)\\\toprule
CoMoSVC-4 &\textbf{4.38 $\pm$ 0.15} 	&\textbf{4.05 $\pm$ 0.19} \\
CoMoSVC-2&4.34 $\pm$ 0.15 	&4.01 $\pm$ 0.19 \\
\textbf{CoMoSVC-1}&4.27 $\pm$ 0.16 	&4.00 $\pm$ 0.19 \\\bottomrule
\end{tabular}
\end{table}

\subsubsection{Effect of Sampling Steps}
In general, as the number of sampling steps increases, there is a small increment in the metrics presented in Table~\ref{table4}. The slight improvement and minor fluctuation indicates that CoMoSVC already achieves accurate score estimation through only one-step discretization yielding high-quality results.

\section{Conclusion}
In this paper, we propose the CoMoSVC, which is based on the consistency model to achieve high-quality, high-similarity and high-speed SVC. The proposed CoMoSVC is a two-stage model where the first stage encodes the features from the waveform, then the second stage utilizes a student model distilled from a pre-trained teacher model to generate converted audios. The comprehensive subjective and objective evaluations demonstrate the effectiveness of CoMoSVC.



\section*{Acknowledgments}
The research was supported by the Theme-based Research Scheme (T45-205/21-N) and Early Career Scheme (ECS-HKUST22201322), Research Grants Council of Hong Kong.
\bibliographystyle{plainnat}
\bibliography{ust_ref}

\end{document}